# Graphene photodetectors for high-speed optical communications

Thomas Mueller[†], Fengnian Xia[*], and Phaedon Avouris[*]

IBM Thomas J. Watson Research Center, Yorktown Heights, NY 10598, USA

[†]current address: Vienna University of Technology, Institute of Photonics, 1040 Vienna, Austria

[*]e-mail: fxia@us.ibm.com, avouris@us.ibm.com

**Abstract. While silicon has dominated solid-state electronics for more than four decades, a variety of new materials have been introduced into photonics to expand the accessible wavelength range and to improve the performance of photonic devices. For example, gallium-nitride based materials enable the light emission at blue and ultraviolet wavelengths[1], and high index contrast silicon-on-insulator facilitates the realization of ultra dense and CMOS compatible photonic devices[2, 3]. Here, we report the first deployment of graphene[4, 5], a two-dimensional carbon material, as the photo-detection element in a 10 Gbits/s optical data link. In this interdigitated metal-graphene-metal photodetector, an asymmetric metallization scheme is adopted to break the mirror symmetry of the built-in electric-field profile in conventional graphene field-effect-transistor channels[6-9], allowing for efficient photo-detection within the entire area of light illumination. A maximum external photo-responsivity of 6.1 mA·W$^{-1}$ is achieved at 1.55 μm wavelength, a very impressive value given that the material is below one nanometer in thickness. Moreover, owing to the unique band structure and exceptional electronic properties of graphene, high speed photodetectors with an ultra-wide operational wavelength range at least from 300 nm to 6 μm[10, 11] can be realized using this fascinating material.**

Graphene, a single layer of carbon atoms assembled in honeycomb lattice, is under intensive investigation currently by both condensed matter physicists and electronic engineers due to its exceptional electronic properties. Graphene also holds great promise for novel photonic devices[10, 12–16]. Recent photocurrent generation experiments in graphene show a strong photo-response near metal/graphene interfaces with an internal quantum efficiency of 15 to 30% despite of its gapless nature[6, 9]. Moreover, graphene photodetectors can potentially be operational at speeds > 500 GHz[17]. However, the external responsivity (or efficiency) of previously demonstrated graphene photodetectors are relatively small, and whether such photodetectors can be useful in realistic applications is not clear. Here we report a simple vertical incidence metal-graphene-metal (MGM) photodetector with external responsivity of 6.1 mA·W$^{-1}$ at 1.55 μm, representing a 15-fold improvement compared with the previous demostration[17]. Most importantly, such graphene photodetectors were deployed in a 10 Gbits/s optical data link for the first time and error-free detection of the optical bit stream was demonstrated, revealing the great potential of graphene in numerous light detection related applications[18-21].

A schematic view of the MGM photodetector is shown in Fig. 1a. The geometry of this MGM photodetector is similar to that of traditional metal-semiconductor-metal (MSM) detectors[22]. Our MGM photodetectors are fabricated on a highly-resistive silicon wafer (1–10 kΩ·cm) with a 300 nm thick thermal oxide. Flakes of single-, bi-, and tri-layer graphenes are identified[23, 24] and confirmed by Raman spectroscopy, and interdigitated electrodes with 1 μm spacing and 250 nm width are then fabricated using electron beam lithography. One set of fingers is made out of Pd/Au (20/25 nm), and the other of Ti/Au (20/25 nm). The active layer of this specific device is bi-layer graphene, and unless otherwise specified, all results reported here are obtained from this device. The active detector area is 6 × 6 μm. The detector is connected to 80 × 80 μm contact pads, separated by 40 μm.

In graphene field-effect-transistor (FET) photodetectors demonstrated previously[17], the built-in electric-field, which is responsible for the separation of the photo-generated carriers, only exists in a narrow region (≈0.2 μm) right at the electrode/graphene interface where charge transfer between the metal and graphene occurs[6-8]. The absence of a strong electric-field in the bulk graphene sheet, where most of the electron-hole pairs are generated, leads to carrier recombination without contribution to the external photocurrent. In the MGM photodetectors reported here, multiple, interdigitated metal fingers are used, leading to greatly enlarged high E-field light detection region. However, if both electrodes in Fig. 1a consist of the same metal, the built-in electric-field profile in the channel between two neighboring fingers is symmetric[6-9], and the total photocurrent is zero. Here, we demonstrate that an asymmetric metallization scheme can be used to break the mirror symmetry of the built-in potential profile within the channel, allowing for summing up their individual contributions to the total photocurrent.

Fig. 1b shows photocurrent images taken at different gate biases $V_G$ between −60 to 60 V applied to the silicon back-gate (see also the corresponding line scans in Fig. 1c). In these measurements, a focused laser beam (He-Ne laser; 633 nm wavelength) is scanned over a device while the induced short-circuit photocurrent is measured as a function of the position of the laser spot on the device to obtain a spatial image with ~0.6 μm resolution[7]. For further details we refer to the Methods section. At large negative voltages ($V_G$ = -60, -40, -30 V), the images show stripes of alternating positive and negative polarity. The integrated response, shown in Fig. 1d, is close to zero as expected, due to rather symmetric potential profile within the graphene channel. At positive voltages ($V_G$ = 0, +10, +20, +40, +60 V), the polarity has switched, but the integral is still zero. We notice, however, that there exists a transition region ($V_G$ = -20, -15, -10, -5 V), where the photocurrent is predominantly positive. This can also be seen in Fig. 1d in which the integrated response exhibits a peak at $V_G$ = -20 V. The electrical characteristic, shown in the inset, depicts the typical shape observed in graphene devices with a maximum resistance of $R_g$ = 140 Ω at gate bias $V_G$ of -5 V ("Dirac voltage"). At large positive and negative gate voltages, the resistance drops due to electrostatic doping of the graphene sheet. The maximum resistance and photo-response do not occur at the same gate bias[6,7].

In order to clarify the origin of this behavior we studied devices consisting of only one pair of fingers and a long (≈2 μm) channel. This allows us to clearly separate the photocurrent contributions from both electrodes, and in Figs. 1e and f we plot their gate voltage dependence. In the device in Fig. 1e, both electrodes consist of the same metal (Pd). The photocurrents generated on the two electrodes have the same magnitude, but opposite polarity. No matter what gate voltage is applied, the sum of the two contributions is always close to zero (the small deviation from zero arises from imperfections of the device). The flipping of the photocurrent polarity occurs at a gate bias $V_G$ around 20 V before the Dirac point voltage is reached[7, 25], due to the p-doping of the graphene introduced by the Pd contact[25]. A schematic of the band profile is shown in the inset. In the other device (Fig. 1f), the two electrodes are made out of different metals (Pd and Ti; as in the multi-finger devices in Fig. 1a). The doping underneath the two electrodes now is different, and the two contributions flip their polarity at different gate voltages. Consequently, when choosing the right gate voltage, the photocurrents around both electrodes can flow in the same direction, leading to an enhanced photocurrent magnitude. The schematic of the band profile at this maximal photocurrent generation condition is shown in the inset of Fig. 1f.

Following the study of the photocurrent generation mechanism in our MGM photodetectors using a He-Ne laser with high spatial resolution, 1.55 μm light is focused on a large spot which illuminates the whole photodetector (shown in the inset of Fig. 2b) for DC and high frequency photocurrent measurements. The gate bias dependent response, plotted in Fig. 2a, resembles the curve determined from the photocurrent images (Fig. 1d) and we determine $V_G$ = -15 V as the optimum operation voltage of our detector. We keep this value fixed in further measurements. Since the gate bias ($V_G$) needed is inversely proportional to the gate oxide thickness, it can be readily reduced by utilizing thinner gate oxide. Moreover, there is little power consumption associated with this relatively large gate bias due to the negligible gate leakage current. The red curve in Fig. 2b shows the power dependence of the DC photocurrent without applying a bias between the interdigitated fingers. The response is linear up to a total optical incidence power $P_{opt}$ on the entire device of 10 mW, and a photocurrent of 15 μA is observed, corresponding to an external responsivity of 1.5 mA/W. A further increase in the incident power leads to photocurrent saturation and hence reduced external responsivity due to the field screening effect as in conventional p-i-n photodiodes[26].

Fig. 2c shows the relative AC photoresponse, measured using a commercial lightwave component analyzer in combination with a parameter network analyzer. A modulated optical signal at 1550 nm wavelength with an average power of 5 mW was focused onto the detector and the electrical output was measured[17]. A 3-dB bandwidth of 16 GHz is determined. Due to the exceptionally high carrier mobility and high saturation velocity, the bandwidth is not likely limited by the carrier transit time. Instead, it is limited by the RC constant of the device. Compared with our previously measured bandwidth in graphene FET based photodetectors, the bandwidth here is smaller. This is due to the relatively large area of the graphene used here, leading to larger device capacitance which is proportional to the area of the device[17].

In order to demonstrate the feasibility of using such graphene based optical components in realistic optical applications, we deployed this graphene photodetector in a 10 Gbits/s optical data link. In this link, the 10 Gbits/s electrical bit stream from a pseudo-random bit sequence (PRBS) generator with ($2^{31}$-1) pattern length is used to modulate a 1550 nm wavelength light. The generated optical bit stream is then amplified to an output power of 17 dBm using an erbium-doped fiber amplifier (EDFA) and focused onto the graphene detector with zero source-drain bias applied. The output electrical data stream produced by the graphene detector is amplified and fed to an oscilloscope to obtain an "eye diagram". (see Methods section for details). As shown in the inset of Fig. 2c, a completely opened eye pattern at 10 Gbits/s is obtained, indicating that this novel 2-dimensional carbon based material system can indeed be used successfully as a high speed detector in photonic systems.

The external responsivity (or efficiency) can be further improved by applying a bias within the photocurrent generation path. In Fig. 3a we show the drain-source bias $V_B$ dependence of the detector current (gate bias $V_G$ stays fixed at -15 V). The black line shows the dark current $I_d$, which is simply given by Ohm's law: $V_B/R_g$, where $R_g$ is the graphene resistance. Under optical illumination (red and blue curves in Fig. 3a), the total current $I$ is the sum of dark current and photocurrent: $I = I_d + I_{ph}$. $I_{ph}$ increases for positive $V_B$ (applied to the Ti electrode; Pd electrode grounded), while it decreases for $0 > V_B > -50$ mV and changes sign at $V_B < -50$ mV. All curves cross at $V_B = -50$ mV, i.e. for this bias the photocurrent turns to zero regardless of the power of optical illumination. This behavior clearly indicates that the E-field within the graphene channel can be influenced by applying an external voltage. From these measurements we can extract important physical parameters related to different metal-graphene contacts. In Fig. 3b we schematically draw the band profiles for three different biasing situations. The image in the middle shows the situation without applying a bias $V_B$ between source and drain. In this case, due to the different doping introduced by Ti and Pd, a potential drop within the graphene channel occurs. The figure on the right shows the case $V_B > 0$. Here, the external bias enhances the potential profile within the channel and the photocurrent increases. On the contrary, at $V_B = -50$ mV (left figure) the external voltage just offsets the built-in potential profile, making the band-profile flat and the photocurrent zero. We hence estimate that in this graphene device, the height of the potential profile within the channel induced by the doping difference of Ti and Pd is around 50 meV. Further reduction of $V_B$ flips the polarity of the photocurrent.

In the inset of Fig. 3a we show the bias dependence of the external responsivity $S$. As $V_B$ increases, $S$ rises monotonously and saturates at $V_B$ around 0.4 V. A maximum responsivity of $S = 6.1$ mA/W is achieved, which represents a 15-fold improvement compared to our previous demonstration[17]. Another simple way of increasing the external responsivity is to use thicker graphene flakes. We have therefore studied the dependence of $S$ on the number of graphene layers at zero source-drain bias $V_B$. Although there is some variation from device to device probably due to the quality of the metal/graphene contacts, the results in Fig. 3c show a clear trend: The responsivity of single-layer graphene photodetectors is on average only half as large as that of the bi-layer detectors. We explain this enhancement by a two-fold increase of absorption in the bi-layer. Increasing the thickness to three layers, however, does not lead to much improvement anymore, because gating and modification of the potential profile within the graphene channel becomes more difficult as more layers are involved due to screening effects[27].

Despite of a 15-fold improvement in the external responsivity in our current device compared with previously demonstrated simple graphene FET based photodetectors[17], the responsivity of 6.1 mA/W is still much smaller than those of conventional InGaAs photodetectors used at 1.55 μm. This is mainly due to the limited absorption of the vertical incidence light, which is about 5% for bi-layer graphene on silicon substrate with a 300 nm oxide. Integration of such graphene photodetectors with an optical waveguide can lead to almost complete absorption of the light[20]. To demonstrate this we performed numerical simulations of a graphene waveguide photodetector consisting of a 0.55 μm wide by 20 μm long bi-layer graphene layer on top of a 0.55 μm wide by 0.22 μm deep silicon photonic wire waveguide (separated by 10 nm of oxide). After 1.55 μm light propagates in the silicon waveguide for 20 μm, >70 % of the light is absorbed. In this case, the total area of the active graphene photodetector is only 11 μm$^2$, and the capacitance is still small enough for high bandwidth operation. Moreover, other combinations of electrode materials could lead to a larger potential difference within the graphene channel and hence further enhanced photocurrent. The photocurrent can also be increased by decreasing the spacing of the metal fingers. In this case, although the height of the graphene channel potential profile remains unchanged, the E-field will be further enhanced. In addition, the recent demonstration of tunable bandgap opening in bi-layer graphene may enable a truly p-i-n based graphene photodetector and further improveperformance[15].

Besides internal E-field which separates the photo-excited electron-hole pairs, photo-thermoelectric effect (PTE) can also contribute to photocurrent generation[28]. However, in our MGM photodetectors, we attribute the efficient photocurrent generation primarily to the internal E-field because of the high photo-responsivity (or efficiency) and large operational bandwidth of the photodetector. The photo-responsivity of our MGM photodetector can be as high as 6.1 mA/W at room temperature, much larger than the previously reported photo-responsivity due to the photo-thermoelectric effect at the single/bi-layer graphene interface. Moreover, the high bandwidth of the photoresponse reported here also implies that photo-thermoelectric effect is not likely the dominant photocurrent generation mechanism in our photodetector.

In summary, for the first time, we deployed a graphene photodetector in a 10 Gbits/s optical data link at the wavelength of 1.55 μm. Error-free signal detection was realized, demonstrating the great potential of this carbon based 2-dimensional material system in photonics. Although the high frequency photocurrent response experiments described here were performed at a single wavelength (1.55 μm), we also observed strong photo-response at three additional wavelengths: 0.514 μm[8], 0.633 μm[7], and 2.4 μm. The estimated internal photo-generation efficiency at 0.633 μm is between 10 and 25%[7]. In addition, we carried out photo-response measurements at 2.4 μm using a pulsed optical parametric oscillator (OPO) source and again a strong photocurrent response was observed. The unique band structure of graphene can enable an ultra-wide operational wavelength range at least from 300 nm to 6 μm. Such exceptionally wide operational wavelength range and intrinsically high operating speed make graphene photodetectors a promising candidate for a variety of applications such as communications, remote sensing, environmental monitoring, and surveillance[18-21]. Moreover, owing to the excellent electronic/photonic properties of graphene and extensive developments in graphene electronics[29], integrated electronic-photonic systems with extremely wide operational wavelength range using graphene are within reach.

**METHODS**

All the measurements were performed in air at room temperature. In order to study the spatially resolved photo-response we use a setup that consists of a He-Ne laser (632.8 nm wavelength) that is focused to a diffraction-limited spot (≈0.6 µm FWHM diameter) on the device using a 50 × microscope objective. A piezo-electrically driven mirror, mounted before the objective, allows positioning the beam on the sample with high spatial precision. The laser beam is modulated with a mechanical chopper (~1 kHz) and the short-circuit photocurrent signal is detected with a current-preamplifier and a lock-in amplifier. The output of the lock-in is fed into a computer to construct the photocurrent image. The back-gate bias $V_G$ is applied using a HP 4145C semiconductor parameter analyzer.

A second setup is used for high-speed optical characterization at a wavelength of 1550 nm. In this setup, the light is focused with an IR lens to a large spot (≈5 µm FWHM diameter) to illuminate the whole active area of the MGM photodetector. Frequency response characterization is achieved using an Agilent Lightwave Component Analyzer N4375B. The optical fiber output of the LCA (0 dBm) is amplified using a PriTel Inc. erbium-doped fiber amplifier (EDFA), coupled into free space, and focused onto the device. The incident optical power on the sample is adjusted by using neutral density filters. The photocurrent signal is extracted through a microwave probe from GGB Industries and fed into a parameter network analyzer (Agilent E8364C). The frequency response (scattering parameter $S_{21}$) is recorded as the modulation frequency is swept between 10 MHz and 20 GHz. This setup also allows performing eye-diagram measurements at 10 GBits/s data rate. For that purpose, a Centellax TG1B1-A pseudo-random bit sequence generator is used to modulate the light from a 1550 nm laser with a JDS Uniphase Mach-Zehnder modulator. The optical signal is then amplified with the EDFA and focused onto the detector. An RF power amplifier with 15 dB gain and 15 GHz bandwidth is used to amplify the detector output and the eye-diagram is measured with an Agilent 86100A wide-band oscilloscope. The gate bias is also applied using a HP 4145C semiconductor parameter analyzer as mentioned above.

**Acknowledgements**

We thank Bruce A. Ek, Damon B. Farmer, and James J. Bucchignano for technical assistance. We are grateful to Yurii A. Vlasov, William M. J. Green, and Solomon Assefa for lending us part of their equipment. T.M. acknowledges financial support by the Austrian Science Fund FWF (Erwin Schrödinger fellowship J2705-N16).

**Additional information**

Correspondence and requests for materials should be addressed to F.X. (fxia@us.ibm.com) and P.A. (avouris@us.ibm.com).

**Competing financial interests**

The authors declare that they have no competing financial interests.

**Figure captions**

Figure 1 **Photocurrent imaging in metal-graphene-metal (MGM) photodetectors with asymmetric metal contacts with 632.8 nm excitation.** (a) Left: a 3-dimensional schematic view of the MGM photodetector. Right: A scanning electron micrograph of the MGM photodetector. Scale bar, 5 μm. The spacing between the metal fingers is 1 μm and the finger width is 250 nm. (b) Photocurrent images of the MGM photodetectors taken at gate biases of -60, -40, -30, -20, -15, -10, -5, 0, 10, 20, 40, and 60 V, respectively. The excitation spot size is about 0.6 μm in diameter. The source-drain bias was kept to be zero. Scale bar, 5 μm. (c) Photocurrent line scans of the identical MGM photodetector at various gate biases from -60 to 60 V (from top to bottom). The scan is performed along the white dashed line shown in the lower right corner of the Fig. 1b. (d) Integrated photocurrent response as a function of the back gate bias. Inset: resistance between the graphene source-drain metal contacts as a function of the back gate bias. (e) Peak photocurrents around source and drain Pd contacts in a graphene field-effect-transistor with 2 μm long channel length. The sum of the photocurrents on both contacts is plotted in the green curve. Inset: the band profile of the graphene field-effect-transistor. (f) Peak photocurrents around source (Pd) and drain (Ti) contacts in a graphene field-effect-transistor with asymmetric metal contacts and 2 μm long channel length. The sum of the photocurrents on both contacts is plotted in the green curve. Inset: the band profile of the graphene field-effect-transistor with asymmetric metal contacts at a gate bias of around -20 V and the sum of the photocurrents is enhanced.

Figure 2 **Photocurrent generation, high frequency characterization of the MGM photodetector, and operation of the MGM photodetector at 10 Gbits/s data rate with 1.55 μm light excitation.** (a) Photocurrent generated at zero-source drain bias as a function of the back gate bias. The excitation light wavelength is 1.55 μm and the spot size is about 5 μm in diameter. The total incidence power at the MGM photodetector is 5 mW. (b) The total photocurrent generated at a back gate bias of around -15 V as a function of the total incidence power. Photocurrent saturation starts at an incidence power of around 10 dBm. The external responsivity before saturation is 1.5 mA/W. Inset: the schematic of the illumination. (c) Relative photoresponse as a function of the light intensity modulation frequency. 3-dB bandwidth of this MSM photodetector is about 16 GHz. Inset: the receiver eye-diagram obtained using this MSM photodetector. A completely open eye is obtained and error-free optical signal detection is demonstrated.

Figure 3 **Source-drain bias ($V_B$) and graphene thickness dependence of the MGM external responsivity.** (a) Current vs. source-drain bias ($V_B$) with and without light illumination. The excitation wavelength is 1.55 μm. The difference between the colored (blue and red) and black lines is the photocurrent. Inset: the measured external photoresponsivity of the MSM photodetector as a function of the source-drain bias. (b) The graphene band profiles at source-drain biases ($V_B$) of -50 mV, 0, and positive bias, from left to right, respectively. (c) External photoresponsivity vs. the number of graphene layers at zero $V_B$. The dark rectangles represent the average external responsivity for single, bi-, and tri- layer graphene photodetectors.

**Figure 1**

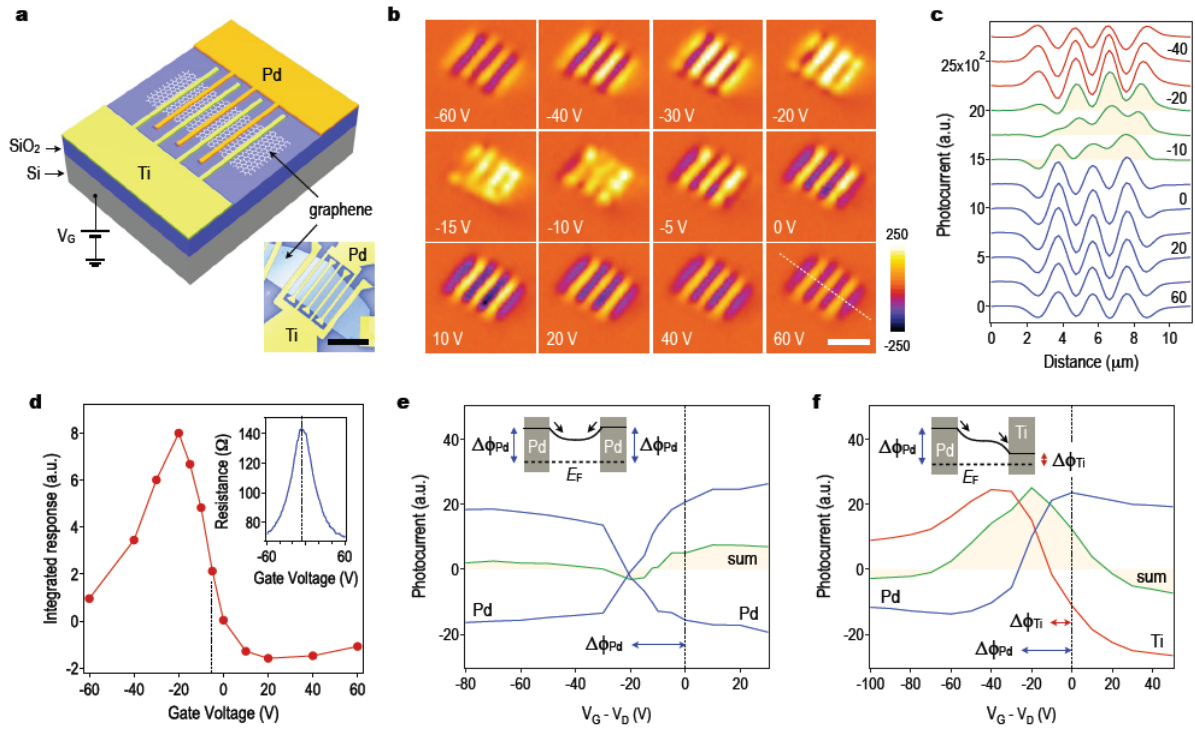

**Figure 2**

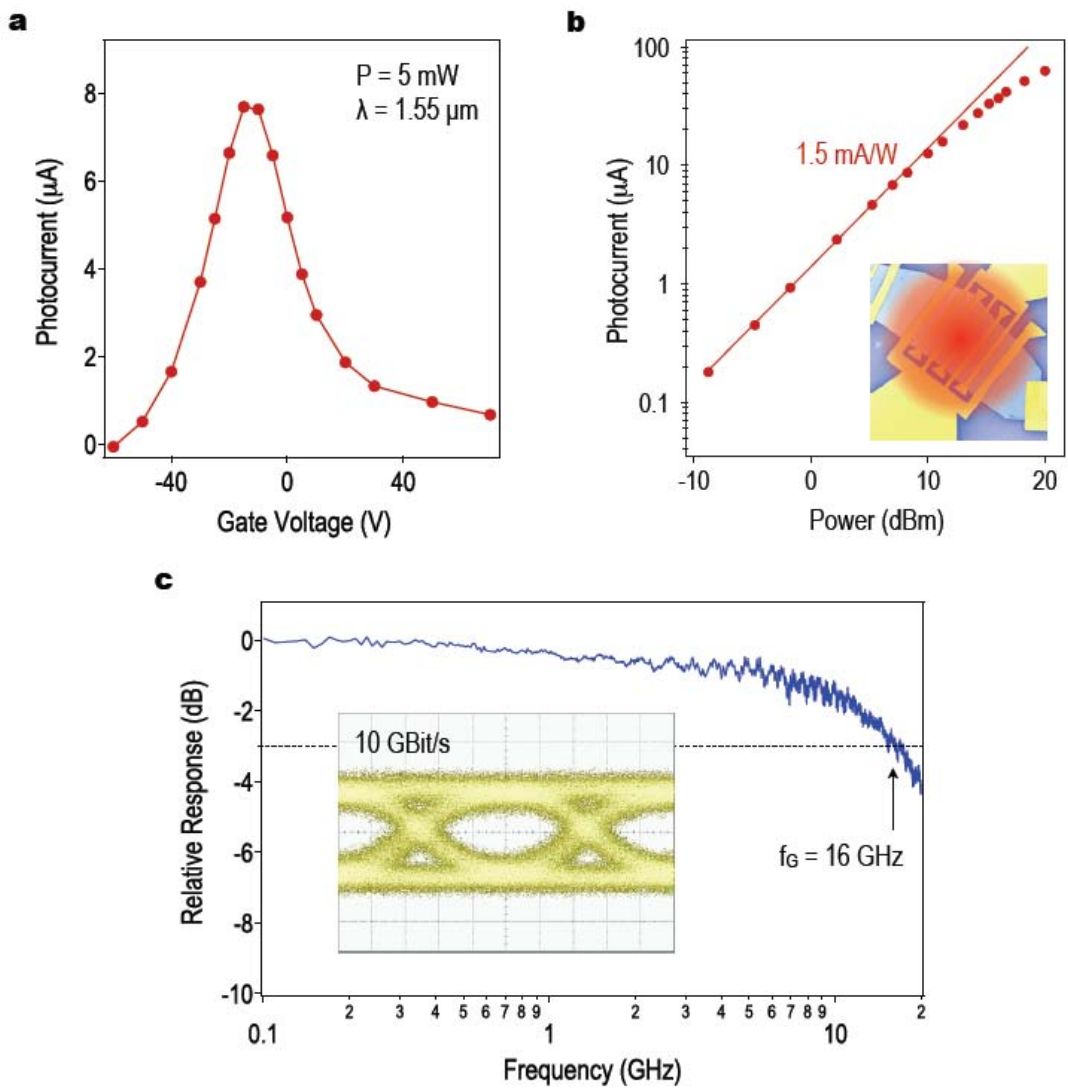

**Figure 3**

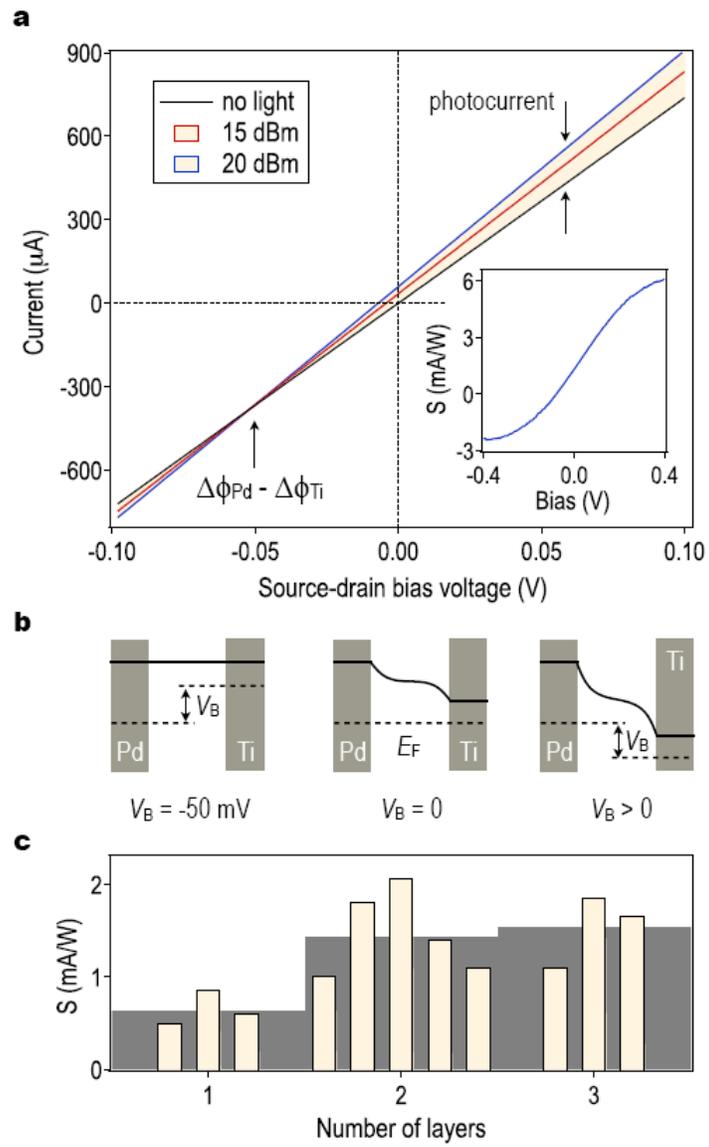